# Design of an efficient Mid-IR light source using As$_2$S$_3$ based highly nonlinear microstructured optical fibers

A. Barh,[1] S. Ghosh,[1] G. P. Agrawal,[2] R. K. Varshney,[1] I. D. Aggarwal,[3] and B. P. Pal.[1*]

[1]Department of Physics, Indian Institute of Technology Delhi, New Delhi, India – 110016.
[2]The Institute of Optics, University of Rochester, Rochester, NY – 14627.
[3]Department of Physics, University of North Carolina, Charlotte, NC-28223.
*Corresponding author: bppal@physics.iitd.ernet.in

**Abstract:** We report on the design of a highly-nonlinear specialty fiber as a mid-infrared light source at 4.3 µm. A meter length of the designed solid-core chalcogenide based index-guided microstructured optical fiber (MOF) with circular air holes has been exploited to translate wavelength via four wave mixing using a thulium-doped fiber laser as the pump with a relatively low peak power of 5 W. A peak gain value of around 37 dB with full width at half maxima (FWHM) less than 3 nm is achieved.

In recent years a strong interest has emerged in developing components and devices capable of operating at wavelengths in the mid-IR range (2 – 25 µm) in view of their potential applications in astronomy, climatology, civil, medical, military, spectroscopy and sensing areas [1]. To design fiber-based devices for these mid-IR applications, chalcogenide glasses are very promising candidate because of their extraordinary linear and nonlinear properties [2-4]. In addition to their chemical durability, such glasses show very good transparency in the mid-IR regime, and their fabrication technology is also well matured [5].

Since chalcogenide glasses exhibit a relatively high nonlinear index co-efficient ($n_2$ ~ 100 times larger than that of conventional silica-fiber) they can enable wavelength conversion from the available near IR sources to the targeted mid- to long-IR range [6]. Additionally, studies on MOFs have shown that the zero dispersion wavelength in such fibers can be tailored to fall within a very broad wavelength range (typically 2 to 11 µm). Moreover, other attractive features like endlessly single mode behavior [7], wide tunability of mode effective area, etc. make MOFs potentially a very suitable candidate to design light sources for mid-IR wavelengths.

For this application-specific fiber design, we choose arsenic sulphide (As$_2$S$_3$) based solid core MOF; reported transmission loss for As$_2$S$_3$-fibers is lowest (< 0.2 dB/m in fibers of lengths ~ 500 m) over the targeted wavelength range among various chalcogenide glasses [5].

In optical fibers several nonlinear phenomena could lead to generation of new wavelengths (s) [2]. Under certain conditions, however, four-wave mixing (FWM) is the dominating candidate for generating new wavelengths, provided a phase-matching condition can be satisfied. For our intended design purpose, we will be dealing with $P_0$ below 5 W, which is considerably lower than the threshold for the onset of stimulated Raman and Brillouin scattering. Therefore in this letter, our aim is to design an As$_2$S$_3$-based solid-core MOF by exploiting degenerate FWM in order to develop a mid-IR source of wavelength in the range of 4 to 5 µm by using commercially available pump sources.

The basic mechanism of FWM is schematically shown in Fig.1 and can be expressed as [8],

$$2\omega_P = \omega_S + \omega_i \qquad (1)$$

where subscripts s, i and p stand for signal, idler, and pump, respectively. Under this interaction two pump photons at frequency $\omega_p$ get converted into a signal photon ($\omega_s$) and an idler photon ($\omega_i$).

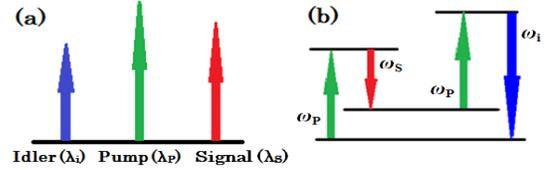

Fig. 1. (Color online) (a) Signal and idler generation using degenerate FWM (single-pump configuration). (b) Energy level diagram of a FWM process.

The equation that governs the propagation of the envelop of signal and idler waves in the fiber, considering dispersion up to second order, can be written as

$$\frac{\partial A_{s,i}}{\partial z} + \beta_{1s,i}\frac{\partial A_{s,i}}{\partial t} + \frac{j}{2}\beta_{2s,i}\frac{\partial^2 A_{s,i}}{\partial t^2} + \frac{1}{2}\alpha_{s,i}A_{s,i}$$
$$= j\gamma\left(|A_{s,i}|^2 + 2|A_{i,s}|^2 + 2P_0\right)A_{s,i} + j\gamma P_0 A_{i,s}^* e^{-j\theta} \qquad (2)$$

where $P_0$ and $\gamma$ stands for the pump power and effective nonlinear coefficient, respectively, $\theta = \kappa z$, and $\kappa$ is the phase mismatch term, which in highly nonlinear fibers is defined as [2]

$$\kappa = \sum_{m=2,4,6\ldots}^{\infty} 2\beta_m(\omega_P)\frac{\Omega_S^m}{m!} + \Delta k_W + 2\gamma P_0 \qquad (3)$$

Here $\Omega_S$ is the frequency shift due to FWM from pump frequency, $\beta_m$ is the $m^{th}$ order dispersion coefficient of the fiber, and $\Delta k_W$ is the phase mismatch due to the wave-guide contribution. We can neglect $\Delta k_W$ for single-mode fibers. To achieve the maximum frequency shift through the FWM process, $\kappa$ should be zero. Thus considering up to fourth order dispersion terms in the expression of $\kappa$, the best route to achieve the phase matching criteria for FWM would be to tailor the dispersion curve through fiber design so as to bring the zero dispersion wavelength ($\lambda_{ZD}$) very close to $\lambda_P$. Our analytical estimates show that maximization of $\Omega_S$ necessitates simultaneous choice of a small positive value of $\beta_2$ and a large negative value of $\beta_4$. This choice leads to the generation of one IR and one mid-IR side band (see Fig. 1). This maximum frequency shift ($\Omega_S$) can be expressed as

$$\Omega_S^2 = \frac{6}{|\beta_4|}\left(\sqrt{\beta_2^2 + \frac{2|\beta_4|\gamma P_0}{3}} + \beta_2\right) \quad (4)$$

To maintain the value of $\beta_4$ negative, the dispersion profile should be flat, which requires the hole-diameter to pitch ratio ($d/\Lambda$) of the MOF to be as large as possible. At the same time, we need to maintain single-mode operation for the pump as well as for the generated signal. For a solid core MOF, if the value of $d/\Lambda$ is less than 0.45, the fiber shows "endlessly single mode" behavior [7]. However, it is very difficult to achieve simultaneously a negative $\beta_4$ value for this particular value of $d/\Lambda$; additionally, to enhance the $\gamma$ value and to minimize the confinement loss at the mid-IR $\lambda_S$, effective mode area ($A_{eff}$) should be as low as possible.

In our design we have balanced all the above mentioned criteria by choosing an optimized set of fiber parameters and by introducing a different size for air-holes in the second cladding ring (radius of air holes is denoted as $r_2$) (see Fig. 2) embedded within the $As_2S_3$ matrix. Here $r$ is the radius of air holes in the remaining rings in the holey cladding. All simulations were carried out by considering a MOF with five cladding rings of air holes as constituting the cladding.

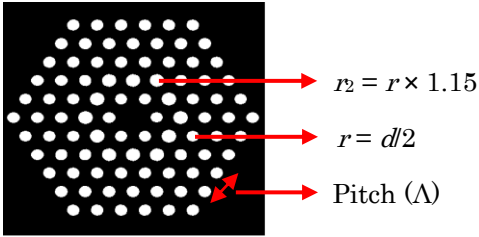

Fig. 2. (Color online) Cross sectional view of the designed MOF with a cladding consisting of 5 rings of air holes (white circles) embedded in the $As_2S_3$ matrix (black background).

To calculate dispersion characteristics of the MOF, wavelength dependence of the linear refractive index [$n(\lambda)$] of the $As_2S_3$ glass has been incorporated through the Sellmeier formula [9]

$$n^2(\lambda) = \sum_j \frac{A_j \lambda^2}{\lambda^2 - \lambda_j^2} \quad (5)$$

where, Sellmeier coefficients $A_1 = 1.898368$, $A_2 = 1.922298$, $A_3 = 0.8765138$, $A_4 = 0.118878$, $A_5 = 0.956998$ and $\lambda_1 = 0.15$µm, $\lambda_2 = 0.25$µm, $\lambda_3 = 0.35$µm, $\lambda_4 = 0.45$µm, $\lambda_5 = 27.3861$µm and $\lambda$ is in µm. The nonlinear index coefficient of $As_2S_3$ is $n_2 = 4.2 \times 10^{-18}$ m²/W. We have chosen the commercially available thulium doped fiber laser at wavelength $\lambda_P = 2.04$ µm as the required pump.

During optimization of the MOF, a strong interplay was evident among $\Omega_S$, parametric gain ($G_S$) and confinement loss ($\alpha_c$) with variations of the parameter $r_2$. After optimization, a high gain of about 37 dB with input pump power of 5 W from one meter of our designed fiber has been achieved for $d/\Lambda = 0.55$, $\Lambda = 1.6$ µm and $r_2 = r \times 1.15$. The dispersion spectrum of the fiber is shown in Fig. 3 from where it is clearly evident that the achieved $\lambda_{ZD}$ at 2.105 µm is very close to $\lambda_P$. The values of $\beta_2$ and $\beta_4$, respectively were 32.68 ps²/km and $-1.6 \times 10^{-3}$ ps⁴/km near pump wavelength.

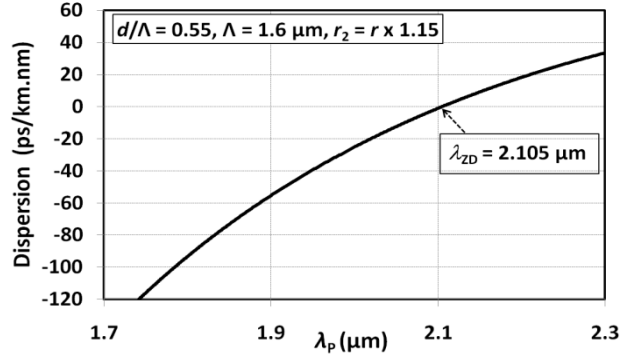

Fig. 3. Calculated dispersion of the designed MOF as a function of wavelength.

The trend of the generated signal and idler wavelengths as a function of different pump wavelengths for a peak pump power of 5 W is shown in Fig. 4. The FWM Gain as a function of $\lambda_S$ is shown in Fig. 5. From Fig. 4 and 5, it can be seen that, for the chosen pump wavelength at 2.04 µm, the signal is generated at 4.34 µm with peak gain value of 37 dB. Moreover, despite the high gain, the gain spectrum is extremely narrow (FWHM ~ 2.6 nm).

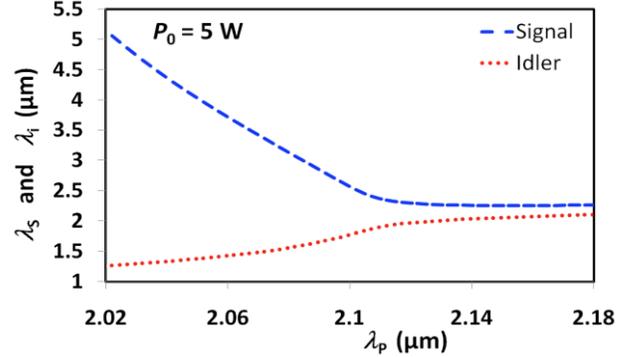

Fig. 4. (Color online) FWM phase-matching curve for designed MOF.

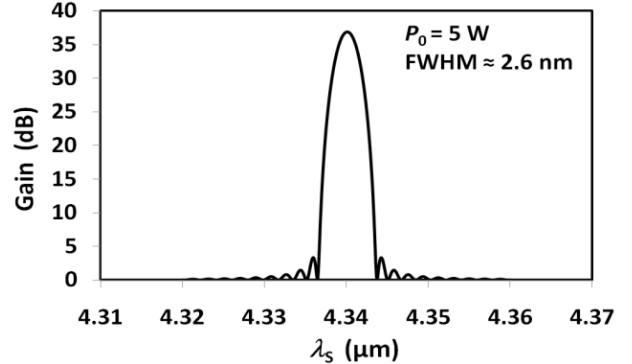

Fig. 5. Gain spectrum of generated signal for $\lambda_P = 2.04$ µm. FWHM is less than 3 nm.

The transverse field pattern of the generated signal at this $\lambda_S$ is shown in Fig. 6. The estimated confinement loss ($\alpha_c$) of the fiber mode at this $\lambda_S$ is around 1.25 dB/m, which is quite favorable to use it as a mid-IR light source. We may mention that by tuning the input pump power ($P_0$) we can further tune the $\lambda_S$ as well as gain ($G_S$) and $\alpha_c$, as shown in Table 1.

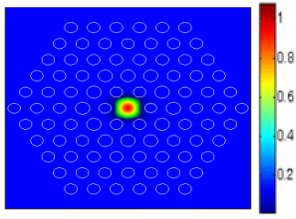

Fig. 6. (Color online) Transverse field distribution of the fiber mode at the generated $\lambda_S$.

Table 1. Comparison of $\lambda_S$, $G_S$, and $\alpha_c$ for three different $P_0$

|  | $P_0 = 1$ W | $P_0 = 5$ W | $P_0 = 10$ W |
| --- | --- | --- | --- |
| $\lambda_S$ (μm) | 4.33 | 4.34 | 4.35 |
| $G_S$ (dB) | 3.7 | 37 | 80 |
| $\alpha_c$ (dB/m) | 1.19 | 1.25 | 1.33 |

The variations of $\lambda_S$, $G_S$, and $\alpha_c$ with the hole size ratio $r_2/r$ is shown in Figs. 7, 8, and 9, respectively. Fig. 7 clearly indicates that as the ratio $r_2/r$ decreases, the wavelength/frequency shift increases. However, this shift is almost independent of pump power. On the other hand, the gain decreases with the decrease of $r_2/r$ (see, Fig. 8). Similar to wavelength shift, $\alpha_c$ also increases with the decrease of $r_2/r$ (see, Fig. 9).

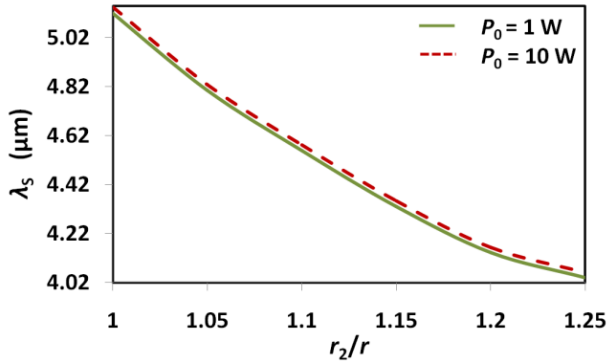

Fig. 7. (Color online) Variation of generated signal wavelength with the ratio $r_2/r$ for two different $P_0$.

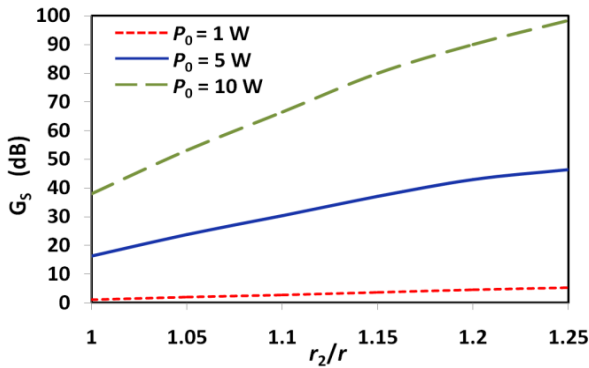

Fig. 8. (Color online) Variation of gain at the generated signal wavelength with the ratio $r_2/r$ for three different $P_0$.

A magnified view of $\alpha_c$ with $r_2/r$ around its optimized value is shown in the inset of Fig. 9. One can appreciate from this figure that the confinement loss is quite high (≈ 120 dB/m) for $r_2 = r$ but around our proposed design parameter ($r_2/r = 1.15$) the loss is significantly low (1.25 dB/m) and it is almost independent of pump power.

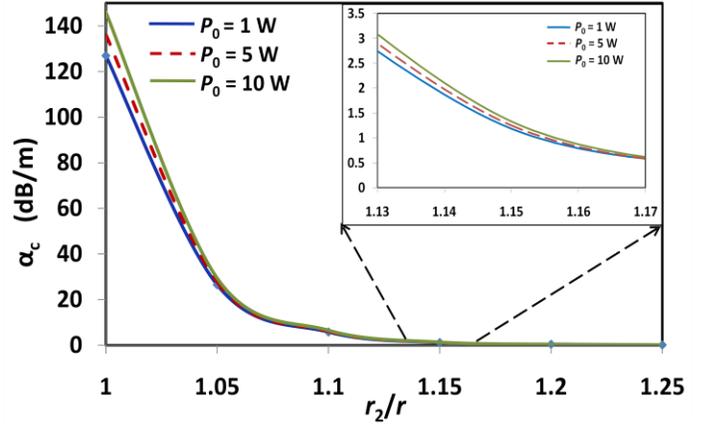

Fig. 9. (Color online) Variation of confinement loss with $r_2/r$ for three different $P_0$.

We may mention that, for the first time to the best of our knowledge, a parametric gain as high as of 37 dB has been achieved with a moderate peak pump power of just 5 W by pumping at 2.04 μm. Additionally, narrow bandwidth (2.6 nm) and very low confinement loss (1.25 dB/m) at generated signal ($\lambda_S$) will make it suitable as an all-fiber monochromatic mid-IR light source for the mid-IR spectroscopy, astronomy and defense applications matching the second low loss transparency window of the terrestrial atmosphere.

This work relates to Department of the Navy Grant N62909-10-1-7141 issued by Office of Naval Research Global. The United States Government has royalty-free license throughout the world in all copyrightable material contained herein.